\documentclass[aps,prl,superscriptaddress,showpacs,twocolumn]{revtex4}

\usepackage{amssymb}
\usepackage{amsmath}
\usepackage{graphicx}

\begin{document}

\title{Spin Accumulation and Spin Relaxation in a Large Open Quantum Dot}
\author{E.~J.~Koop}
\author{B.~J.~van~Wees}
\affiliation{Physics of Nanodevices Group, Zernike Institute for
Advanced Materials, University of Groningen, Nijenborgh 4, NL-9747AG
Groningen, The Netherlands}
\author{D.~Reuter}
\author{A.~D.~Wieck}
\affiliation{Angewandte Festk\"{o}rperphysik, Ruhr-Universit\"{a}t Bochum,
D-44780 Bochum, Germany}
\author{C.~H.~van~der~Wal}
\affiliation{Physics of Nanodevices Group, Zernike Institute for Advanced
Materials, University of Groningen, Nijenborgh 4, NL-9747AG Groningen,
The Netherlands}
\date{\today}

\begin{abstract}

We report electronic control and measurement of an
imbalance between spin-up and spin-down electrons in
micron-scale open quantum dots. Spin injection and
detection was achieved with quantum point contacts tuned to
have spin-selective transport, with four contacts per dot
for realizing a non-local spin-valve circuit. This provides
an interesting system for studies of spintronic effects
since the contacts to reservoirs can be controlled and
characterized with high accuracy. We show how this can be
used to extract in a single measurement the relaxation time
for electron spins inside a ballistic dot ($\tau_{sf
}\approx 300 \; {\rm ps}$) and the degree of spin
polarization of the contacts ($P \approx 0.8$).

\end{abstract}



\maketitle


The ability to control and detect the average spin
orientation of electron ensembles in non-magnetic
conductors lies at the heart of spintronic functionalities
\cite{6Awschalom2007}. We report here electronic control
and detection of spin accumulation --an imbalance between
the chemical potential of spin-up and spin-down electrons--
in a large ballistic quantum dot in a GaAs heterostructure.
We use quantum point contacts (QPCs) to operate a
four-terminal quantum dot system, which is suited for
realizing a non-local spin-valve circuit
\cite{6Jedema2001}. Before, such spin-valve circuits were
realized with ferromagnetic contacts on various
non-magnetic conductors
\cite{6Jedema2001,6Lou2007,6Tombros2007}, but for these
systems it is hard to characterize the contact properties.
An interesting aspect of our spintronic system is that it
is realized with ultra-clean non-magnetic materials, while
each spin-selective mode in the contacts can be controlled
individually. We demonstrate that this can be exploited to
measure and unravel for a single device the spin relaxation
rate inside the dot, contributions to spin relaxation from
coupling the dot to reservoirs, and the degree of
polarization for spin-selective transport in the contacts.
Thus, we report here the spin relaxation time for two
different confinement geometries. Chaotic scattering inside
such ballistic cavities can result in a spin relaxation
mechanism that differs from that of bulk materials and very
small few-electron quantum dots \cite{6Koop2008}, but its
full understanding is still a challenge to the community
\cite{6Beenakker2006}.

\begin{figure}
\includegraphics[width=1.0\columnwidth]{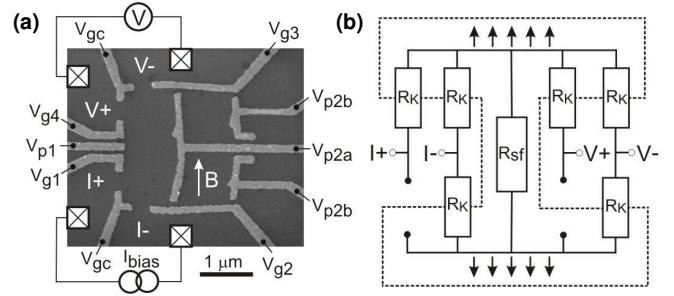} \caption{(a) Electron
microscope image of the device, with labels for current and voltage
contacts, and depletion gates $V_{gi}$ and $V_{pi}$. Gate $V_{p1}$
is a shape distorting gate. Fully switching gate $V_{p2a}$ or gates
$V_{p2b}$ on or off sets the overall size of the dot, but fine
tuning these gates is also used for controlling small shape
distortions. (b) Resistor model for the most typical experiment (see
text), for the case of ideal spin polarization of the contacts to
the $I+$ and $V+$ reservoirs. The spin-up (top) and spin-down
(bottom) populations inside the dot are contained within the dashed
line. The spin-flip resistance $R_{sf}$ represents spin relaxation
inside the dot. } \label{6DeviceAndRmodel}
\end{figure}

Figure~\ref{6DeviceAndRmodel}a presents our device.
Depletion gates on a heterostructure with a two-dimensional
electron gas (2DEG) below the surface are used to define
the four-terminal dot. QPCs are operated as spin-selective
contacts, using that the subbands that carry the ballistic
transport can be Zeeman split with a strong in-plane
magnetic field, and that these modes can be opened up one
by one by tuning gate voltages \cite{6Potok2002,6Koop2007}.
The conductance of QPCs then increases in steps, with
plateaus at $N e^2/h$, where $N$ the number of open modes.
For odd (even) $N$ the last opened mode carries only
spin-up (spin-down). For the most typical form of our
experiment we tune to the following setting. The QPC to the
$I+$ reservoir has a single open mode, which is only
available for spin-up electrons, while the $I-$ QPC is
tuned to carry one mode for spin-up and one for spin-down,
and we apply here a current $I_{bias}$. The contact
resistance for electrons entering the dot via $I-$ is equal
for spin-up and spin-down, while the current that leaves
the dot carries only spin-up. Consequently, the chemical
potential for spin-down electrons inside the dot will
become higher than that for spin-up, up to a level that is
limited by spin relaxation. This difference in chemical
potential $\Delta \mu_{\uparrow \downarrow}$ can be
measured as a voltage: with the $V+$ QPC tuned to have only
one open mode for spin-up and the $V-$ QPC tuned to have
one open mode for spin-up and one for spin-down, the
voltage is $V = \Delta \mu_{\uparrow \downarrow} /2e$,
which is for linear response expressed as a non-local
resistance $R_{nl}=V/I_{bias}$.


The resistor model in Fig.~\ref{6DeviceAndRmodel}b is
useful for analyzing how spin-relaxation mechanisms
influence the measured signal in the above experiment. Each
open mode for spin-up in a QPC is modeled as a resistor
with value $R_{K}=h/e^2$ to the spin-up population in the
dot, and similar for spin-down (we assume first perfect
polarization of QPCs tuned to be spin selective).
Spin-relaxation \textit{inside} the dot is modeled as a
resistor $R_{sf}$ that carries a current from the spin-up
to the spin-down population. Figure~1b illustrates that the
contacts to the $I-$ and the $V-$ reservoir provide
additional current paths for relaxation parallel to
$R_{sf}$ (spins rapidly mix in reservoirs, and reservoirs
always have zero spin-accumulation). This mechanism for
spin relaxation \textit{outside} the dot causes that in the
limit of $R_{sf} \rightarrow \infty$ (no relaxation inside
the dot), $R_{nl}$ is limited to $R_K / 4$. The voltage
that is driving the relaxation inside the dot is $\Delta
\mu_{\uparrow \downarrow}/e$, while the current through
$R_{sf}$ is $I_{sf} = e \Delta \mu_{\uparrow \downarrow} /
2 \Delta_{m} \tau_{sf}$, such that the spin-flip time
$\tau_{sf}$ dictates $R_{sf}$ according to $R_{sf}=  2
\tau_{sf} \Delta_{m} /e^2$ \cite{6Johnson1993}. Here
$\Delta_{m} = 2 \pi \hbar^2 / m^{*} A$ is the mean energy
spacing between spin-degenerate levels in a dot of area
$A$. Consequently, measuring $R_{nl}$ and deriving $R_{sf}$
from its value can be used for determining $\tau_{sf}$.
While this resistor model does not account for various
mesoscopic effects that occur in ballistic chaotic quantum
dot systems, a theoretical study of an equivalent
two-terminal spintronic dot \cite{6Beenakker2006} showed
that it is valid in the regime that applies to our
experiment (no influence of weak-localization and Coulomb
blockade effects), and we indeed find that it is consistent
with the measured spin signals that we report.


The dot was realized in a GaAs/Al$_{0.32}$Ga$_{0.68}$As
heterostructure with the 2DEG at 114 nm depth. At 4.2 K,
the mobility was $\mu = 159 \; {\rm m^{2}/Vs }$ and the
electron density $n_{s} = (1.5 \pm 0.1) \cdot 10^{15} \;
{\rm m^{-2}}$. For gates we used electron-beam lithography
and lift-off techniques, and deposition of 15 nm of Au on a
Ti sticking layer. The reservoirs were connected to wiring
via Ohmic contacts, which were realized by annealing
Au/Ge/Ni from the surface. All measurements were performed
in a dilution refrigerator at an effective electron
temperature $ T_{eff} \approx 100 \; {\rm mK}$. For
measuring $R_{nl}$ we used lock-in techniques at 11 Hz with
a current bias, where we made sure that the associated bias
voltage $V_{bias} \leq 10 \; {\rm \mu V}$. We carefully
checked that RC-effects did not influence $R_{nl}$ results.
We used the T-shaped gate $V_{p2a}$ or pair of gates
$V_{p2b}$ for setting the \textit{overall size} of the dot
(not to be confused with tuning small shape distortions for
averaging out fluctuations, see below) at either an area of
$1.2 \; \rm{ \mu m^2 }$ or $2.9 \; \rm{ \mu m^2 }$
(accounting for a depletion width of $\sim 150 \; \rm{nm}$
around the gates).

\begin{figure}
\includegraphics[width=1.0\columnwidth]{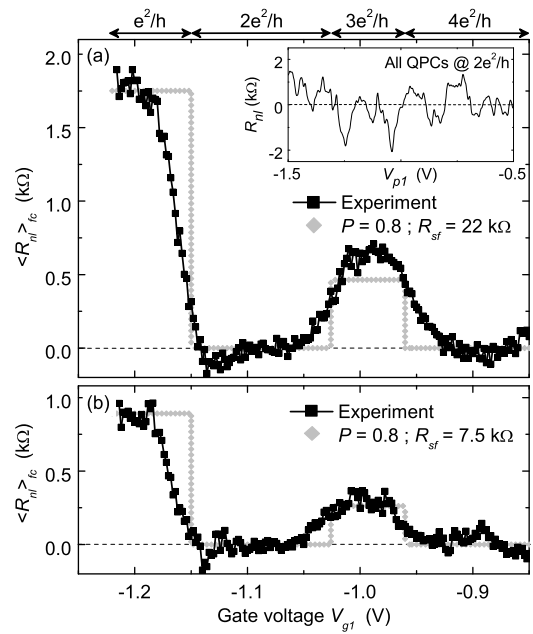}
\caption{(a) Non-local resistance results $\langle R_{nl}\rangle _{fc}$
as a function of
gate voltage $V_{g1}$ (controlling the number of open modes in the
$I+$ QPC, corresponding conductance plateaus are
indicated at the top axis) for
the dot with area 1.2 $\mu m^2$ (a) , and 2.9 $\mu
m^2$ (b), measured at $B = +8.5 \; \rm{T}$.
Gray lines show $R_{nl}$ values from the resistor model,
with the spin-flip resistance $R_{sf}$ and polarization $P$ as in
the figure labels. The inset in (a) shows fluctuations of $R_{nl}$
as a function of shape gate $V_{p1}$ with all QPCs at a
conductance of $2e^2/h$.} \label{6RnlvsIplus}
\end{figure}

Before presenting measurements of spin accumulation, we
discuss two effects that make this experiment in practice
less straight forward than in the above description.
Quantum fluctuations in $R_{nl}$ due to electron
interference inside the dot \cite{6Alhassid2000} have an
amplitude that is comparable to the spin signal
\cite{6Lerescu2008}, and $R_{nl}$ can only be studied as a
spin signal after averaging over a large number of
fluctuations. The inset of Fig.~\ref{6RnlvsIplus}a shows
such fluctuations in $R_{nl}$ as a function of the voltage
on $V_{p1}$, which causes a small shape distortion of the
dot. We discuss results as $\langle R_{nl} \rangle $ when
presenting the average of 200 independent $R_{nl}$
fluctuations, from sweeping with two different
shape-distorting gates. Cross talk effects between gates
were carefully mapped out and compensated for keeping the
QPCs at their desired set points \cite{6Stewart1999}.


A second effect which, besides spin accumulation, may
result in strong $R_{nl}$ values is electron focusing
\cite{6Lerescu2008}. Our sample was mounted with its plane
at $0.73^{\circ}$ with respect to the direction of the
total magnetic field $B$. Consequently, there is a small
perpendicular field $B_{\bot}$ and the associated electron
cyclotron diameter equals the $I+$ to $V+$ contact distance
(Fig.~\ref{6DeviceAndRmodel}a) at $B = \pm 6 \; \rm{T}$. We
will mainly present results measured at $B = +8.5 \;
\rm{T}$, for which we found that focusing only weakly
influences $\langle R_{nl} \rangle$ results. Further, we
use that we can subtract a background contribution to
$\langle R_{nl} \rangle$ from focusing (discussed below),
and we present results where this is applied as $\langle
R_{nl} \rangle _{fc} $. $B_{\bot}$ also breaks
time-reversal symmetry (suppressing weak localization) when
$|B| > 0.2 \; \rm{T}$.


Figure~\ref{6RnlvsIplus}a presents $\langle R_{nl} \rangle
_{fc} $ as a function of the number of open modes in the
$I+$ contact (tuned by $V_{g1}$), while the other QPCs are
tuned as in Fig.~\ref{6DeviceAndRmodel}b. On the left on
this $V_{g1}$ axis, the $I+$ QPC carries only one spin-up
mode (conductance $G_{I+}$ tuned to the $e^2/h$ plateau,
see also top axis). Here $\langle R_{nl} \rangle _{fc}
\approx 1.8 \; \rm{k \Omega}$. Tuning $V_{g1}$ to more
positive values first adds an open spin-down mode to the
$I+$ QPC ($G_{I+}$ at $2 e^2/h$), such that it is no longer
spin selective and $\langle R_{nl} \rangle _{fc} $ drops
here indeed to values near zero. Further opening of the
$I+$ QPC tunes it to have two spin-up modes in parallel
with one-spin down mode ($G_{I+}=3 e^2/h$). This causes
again a situation with more spin-up than spin-down current
in the $I+$ QPC, but less distinct than before and here
$\langle R_{nl} \rangle _{fc} $ shows again a clear
positive signal. Then, it drops to zero once more when the
next spin-down mode is opened in the QPC. We obtain
nominally the same results when the role of the current and
voltage contacts is exchanged. Further,
Fig.~\ref{6RnlvsIplus}b shows that the large dot shows the
same behavior, but with lower $\langle R_{nl} \rangle _{fc}
$ values. This agrees with a lower value for $R_{sf}$ for
the large dot. From these measurement we can conclude that
$\langle R_{nl} \rangle _{fc} $ is a signal that is
proportional to the spin accumulation $\Delta \mu_{\uparrow
\downarrow}$ in the dot.


\begin{figure}
\includegraphics[width=1.0\columnwidth]{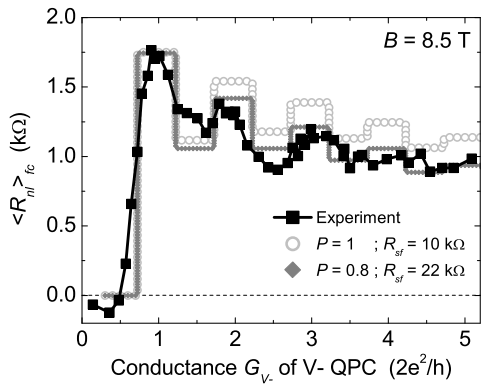}
\caption{Averaged non-local resistance $\langle R_{nl} \rangle
_{fc}$ as a function of the conductance $G_{V-}$ of the V- QPC, for
$A= 1.2 \; \rm{\mu m^2}$. Gray lines show $R_{nl}$ values
from the resistor model, with the spin-flip resistance $R_{sf}$ and
polarization $P$ as labeled.} \label{6RnlvsVmin}
\end{figure}


Figure~\ref{6RnlvsVmin} shows results from a similar
experiment on the small dot (but also here the large dot
showed the same behavior). Now $\langle R_{nl} \rangle
_{fc} $ is measured as a function of the number of open
modes in the $V-$ QPC (tuned by $V_{g3}$), while all other
QPCs are again tuned as in Fig.~\ref{6DeviceAndRmodel}b.
Here we observe a signal close to zero when the $V-$ QPC
carries only one spin-up mode ($G_{V-}=e^2/h$) since it
then probes the same chemical potential as the $V+$ QPC.
Opening it to $G_{V-}=2e^2/h$ immediately results in a
strong signal. Further opening this QPC then causes the
signal to go up and down, qualitatively in reasonable
agreement with the resistor model that assumes perfect
polarization ($P=1$) of each spin-selective mode in a QPC
(see theory traces in Fig.~\ref{6RnlvsVmin}, these go up
and down in a step-like manner since we assume sharp
transitions between conductance plateaus). However, with
quantitative agreement at $G_{V-}=2e^2/h$ (for $R_{sf} = 10
\; \rm{k \Omega}$), this model with $P=1$ shows an average
slope down with increasing $G_{V-}$ that is too weak.
Instead, we find that the resistor model can show
quantitative agreement over the full $G_{V-}$ range (and
with the results in Fig.~\ref{6RnlvsIplus}) when we account
for imperfect spin polarization of QPCs.

We model imperfect polarization in the resistor model as follows. We
assume it only plays a role for QPCs set to a conductance of $N
e^2/h$, with $N$ an odd integer (because the energy spacing between
pairs of Zeeman-split subbands is large \cite{6Koop2007}).
Spin-selective transport is then only due to the highest pair of
subbands that contributes to transport, and \textit{we define the
polarization $P$ only with respect to this pair}. This pair of
subbands is then modeled as a resistor $R_{\uparrow} = 2 R_K /
(1+P)$ to the spin-up population in the dot, and a resistor
$R_{\downarrow} = 2 R_K/(1-P)$ to the spin-down population, which
corresponds to $P = ( R_{\downarrow} - R_{\uparrow} )/(
R_{\downarrow} + R_{\uparrow} )$. This provides a simple model for
$R_{nl}$ with only $R_{sf}$ and $P$ as fitting parameters if we
assume that all spin-selective QPCs and QPC settings can be modeled
with a single $P$ value. We find then a good fit to all the data in
Figs.~\ref{6RnlvsIplus} and \ref{6RnlvsVmin} for $P=0.8 \pm 0.1$,
with $R_{sf} = 22 \pm 3 \; \rm{k \Omega}$ for the small dot and
$R_{sf}= 7.5 \pm 1 \; \rm{k \Omega}$ for the large dot. In
Fig.~\ref{6RnlvsIplus}a at $G_{I+}=3e^2/h$, the experimental results
are higher than the plotted model values. However, this turns into
the opposite situation when using results obtained with the current
and voltage QPCs exchanged. This indicates that $P$ does not have
exactly the same value for all QPCs and QPC settings. There is,
however, always agreement with the model when accounting for the
error bars of $P$ and $R_{sf}$.

The values of $R_{sf}$ correspond to $\tau_{sf}= 295 \pm 40
\; \rm{ps}$ for the small dot and $\tau_{sf}= 245 \pm 35 \;
\rm{ps}$ for the large dot. In our type of system spin
relaxation in the dot is probably dominated by Rashba and
Dresselhaus spin-orbit coupling. How this mechanism results
in a certain value for $\tau_{sf}$ then depends on the
ballistic scattering rate at the edge of the dot. We
performed numerical simulations of this mechanism, which
yield that relaxation times indeed depend on the size of
the dot, with typical values near 300 ps \cite{6Koop2008}.
In our experiment, however, the error bars for $\tau_{sf}$
are too large for studying this dependence on the shape of
our dots, but our method is suited exploring this topic in
future work.


\begin{figure}[t]
\includegraphics[width=1.0\columnwidth]{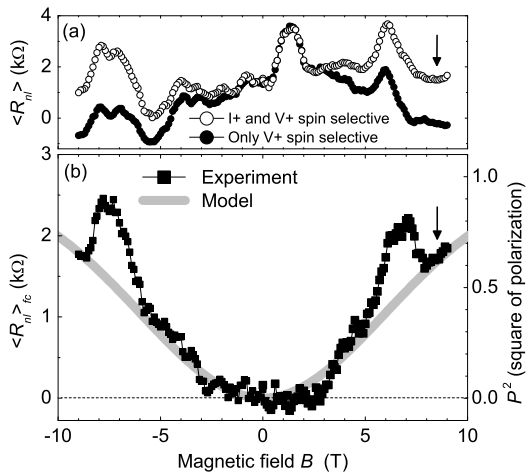}
\caption{(a) Averaged non-local resistance $\langle R_{nl}
\rangle $ as a function of $B$, for $I+$ and $V+$ at the
$e^2/h$ spin-polarized conductance plateau (open symbols),
and for $I+$ at $2e^2/h$ (not spin selective) and only $V+$
at $e^2/h$ (closed symbols). The $I-$ and $V-$ QPCs are at
$2e^2/h$, $A = 1.2 \; \rm{\mu m^2}$. The difference in
$\langle R_{nl} \rangle $ for the traces in (a) defines the
focusing corrected non-local resistance $\langle R_{nl}
\rangle _{fc}$, shown in (b). The gray line in (b) is a fit
of the model where the polarization $P$ of QPCs (right
axis) increases with Zeeman-splitting (see text). Arrows
indicate $B$ that was applied for measuring the data of
Figs. \ref{6RnlvsIplus}, \ref{6RnlvsVmin}.} \label{6RnlvsB}
\end{figure}


Figure~\ref{6RnlvsB} shows how focusing affects $\langle
R_{nl} \rangle $ and $\langle R_{nl} \rangle _{fc}$. For
QPCs tuned as in Fig.~\ref{6DeviceAndRmodel}b the signal
from spin accumulation drops to zero if either the $I+$ or
the $V+$ QPC is tuned from $e^2/h$ to $2 e^2/h$ (no longer
spin selective). However, when sweeping $B$ we also measure
large positive and negative $\langle R_{nl} \rangle$ values
when the $I+$ QPC, the $V+$ QPC or both are at $2 e^2/h$.
For these three settings we observed $\langle R_{nl}
\rangle $ traces that are nominally the same (black symbols
in Fig.~\ref{6RnlvsB}a). The peaked structure is due to
electron focusing effects \cite{6Potok2002,6Lerescu2008}.
Only the peak at $+6 \; {\rm T}$ corresponds to direct
focusing from the $I+$ into the $V+$ contact without an
intermediate scatter event on the edge of the dot (it has
the right $B$ value and other peaks move to other $B$
values when comparing the small and the large dot). Note,
however, that all $\langle R_{nl} \rangle $ values are
significantly higher when both the $I+$ and $V+$ QPC are
tuned to be spin selective (open symbols in
Fig.~\ref{6RnlvsB}a). This difference between the open and
black symbols defines the quantity $\langle R_{nl} \rangle
_{fc}$ (Fig.~\ref{6RnlvsB}b) and provides a signal that is
mainly due to spin. This $\langle R_{nl} \rangle _{fc}$
data also shows a peaked structure where $\langle R_{nl}
\rangle $ shows strong focusing signals. This agrees with
enhancement of electron focusing signals between
spin-selective QPCs \cite{6Potok2002}.

For interpreting $\langle R_{nl} \rangle _{fc}$ as a
measure for spin accumulation, the experiment must be
performed in a regime with many chaotic scatter events
inside the dot during the electron dwell time. This is
clearly not the case at the focusing peaks in
Fig.~\ref{6RnlvsB}b (at $-7.5 \; {\rm T}$ and $+6 \; {\rm
T}$). We therefore studied spin accumulation at $+8.5 \;
{\rm T}$ where focusing from the $I+$ QPC scatters on the
edge of the dot just before the $V+$ contact and where the
signatures of focusing in $\langle R_{nl} \rangle $ are
small. The agreement between the results of both
Figs.~\ref{6RnlvsIplus} and \ref{6RnlvsVmin}, for both the
small and large dot, and the resistor model supports the
conclusion that these results were obtained in a chaotic
regime.

As a final point we discuss that the degree of polarization
$P=0.8$ is in agreement with independently determined QPC
properties. Steps between conductance plateaus are
broadened by thermal smearing (a very weak contribution for
our QPCs at 100 mK) and due to tunneling and reflection
when the Fermi level $E_F$ is close to the top of the QPC
potential barrier for the mode that is opening. It is
mainly this latter effect that causes $P<1$ in our
experiments. The role of tunneling and reflection in QPC
transport is described with an energy dependent
transmission $T(\epsilon)$ that steps from 0 to 1 when a
QPC mode is opened. We study the effect of this on $P$ by
assuming that $E_F$ is located exactly between the bottoms
of a pair of Zeeman split subbands. For these two subbands
we use $T(\epsilon)_{\uparrow(\downarrow)} = ({\rm erf}(
\alpha(\epsilon - E_F -(+) E_Z/2)  )+1)/2$, a
phenomenological description that agrees with studies of
our QPCs \cite{6Koop2007}. Here $E_Z = g \mu_B B$ is the
Zeeman splitting (for g-factor $g$ and Bohr magneton
$\mu_B$) and $\alpha$ a parameter that sets the width of
the step in $T(\epsilon)$. For $e V_{bias} < k_B T_{eff}$,
the contributions of these two subbands to the QPC
conductance are then $G_{\uparrow(\downarrow)} = (e^2/h)
\int  {\rm d} \epsilon \left(-{\rm d} f/  {\rm d} \epsilon
\right) T(\epsilon)_{\uparrow(\downarrow)}$, where $f$ the
Fermi function. With $P = ( G_{\uparrow} - G_{\downarrow}
)/( G_{\uparrow} + G_{\downarrow} )$ we now calculate how
$P$ increases with $B$ due to an increasing Zeeman
splitting. In the resistor model the dependence of $R_{nl}$
on $P$ is close to $R_{nl} \propto P^2$. We therefore plot
$P^2$ in Fig.~\ref{6RnlvsB}b (gray line, with scaling of
the right axis such that it overlaps with the experimental
results) for parameters that give the consistent result
$P=0.8$ at $B=8.5 \; {\rm T}$. For this we use $|g|=0.44$
(as for bulk GaAs) and an $\alpha$ value that is derived
from a full-width-half-max of 0.2 meV for the peak in ${\rm
d} T(\epsilon) / {\rm d} \epsilon$. The latter parameter
agrees with the values 0.20 to 0.35 meV that we found when
characterizing this for our QPCs \cite{6Koop2007}. Notably,
we cannot calculate such a consistent result if we assume
that the many-body effects that we observed in our QPCs
\cite{6Koop2007} enhance the Zeeman splitting (showing for
example  $|g| \approx 1.1$). This indicates that these
effects do not play a role for spin injection and detection
with QPCs, as was also found in Ref.~\cite{6Potok2002}.


We thank A. Lerescu, B. Wolfs and D. Zumb\"{u}hl for useful
discussions, and the Dutch FOM and NWO, and the German DFG-SFB 491
and BMBF nanoQUIT for funding. We have been made aware of related
results by S. M. Frolov \textit{et al.} \cite{6Folk2008} with a
narrow Hall bar and D. M. Zumb\"{u}hl \textit{et al.} with a
two-terminal dot \cite{6Zumbuhl2001}.


\end{document}